\documentstyle[prb,aps]{revtex}

\begin{document}


\title{ QUANTUM TUNNELING IN FERRIMAGNETIC NANOPARTICLES
COUPLED TO A SPIN BATH: AN EFFECTIVE HAMILTONIAN} 

\author{ I.S. Tupitsyn }
\address{Russian Research Center "Kurchatov Institute", 123182 Moscow, 
Russia.}

\maketitle

\vspace{1cm}

\begin{abstract}
An effective Hamiltonian, describing quantum tunneling in {\it 
ferrimagnetic} nanoparticles which includes 
interactions between the electronic spins 
of nanoparticle and microscopic environmental spins (like nuclear spins or 
paramagnetic impurities), is obtained. Two limiting cases, describing 
tunneling in antiferromagnetic and ferromagnetic regimes are
considered, and 
criterion for the transition between the two regimes is found. The validity
of analytic results is verified by the exact diagonalization method. 

\end{abstract}




\begin{center}
{\bf I. INTRODUCTION} 
\end{center}

The quantum tunneling of magnetization (or Neel vector) 
has been attracting a lot of attention during the last years 
from both experimental and theoretical point of view since this 
quantum effect might be seen on macroscopic scale for the total number of 
spins of order $10^4$ and more. Electronic  spins in the magnetic grain 
are coupled to each other by the exchange interaction, forming either a 
``giant spin'' in ferromagnetic case or a ``giant Neel vector'' in 
antiferromagnetic case, and can tunnel coherently between the two 
degenerate states separated by the magnetic anizotropy barrier. The 
theory of coherent tunneling of the 
magnetization vector has been developed in many papers [1] (for a
review see [2] and refs therein). As to the antiferromagnetic particle, the 
tunneling effect in this case may be even stronger than in the ferromagnetic 
one [4] (for a review of modern state of the theory see [3-9]), and of the 
experiment - [10].) It turns out that in both cases interactions with a 
spin bath (like nuclear spins or paramagnetic impurities) play a very 
important role [11,13]. These interactions may be so strong that 
they can completely suppress the quantum coherence [12]. 
For the theoretical analysis of the tunneling dynamics  one usually 
introduces an {\it effective Hamiltonian}, $H_{eff}$, 
operating only in the  low-energy 
subspace of the combined system - ``giant spin'' (or ``giant Neel vector'')
plus the spin bath. This procedure involves truncation
of the original problem to the subspace of the two lowest 
energy levels corresponding to opposite spatial orientations of the order 
parameter, i.e., all transitions from the lowest doublet 
to states with exitation energy $E \buildrel{_>}\over {_\sim} \Omega_o $, 
where $ \Omega_o$ characterizes the small oscillation frequency
of the order parameter near the classical minima, are ignored in 
$H_{eff}$. Earlier $H_{eff}$  for the ``giant spin" was 
derived by the instanton technique in [15]. In this paper I obtain an 
analogous Hamiltonian for the case of {\it ferrimagnetic/antiferromagnetic} 
nanoparticles. To explore the limits of validity of 
the analytical expressions I use the exact diagonalization technique. 

\bigskip

\begin{center}
{\bf II. HAMILTONIAN AND MODEL}
\end{center}


\begin{center}
{\bf A. STARTING MODEL } 
\end{center}

Taking advantage of the known fact that Heisenberg model with 
antiferromagnetic ordering can be reduced in the 
long-wave limit to the non-linear 
O(3) $\sigma$-model (both in 1D and 2D cases) [3,5,7,9,16], we introduce an 
effective two-sublattice model with strong exchange interaction between 
sublattices, the Lagrangian being analogous to that of O(3) $\sigma$-model 
(see, for example, [3,4,8,18]). In addition to the just mentioned 
works we include coupling between the sublattice spins 
$\vec{S}_1$ and $\vec{S}_2$ and environmental spins 
($\mid \vec{\sigma}_i \mid={1 \over 2}$) 
in a weak magnetic field $\vec{H}_o$. Since coupling to a single
environmental spin is typically rather small 
and can be treated perturbatively,
it is sufficient, without 
loss of generality, to consider here just one such spin coupled, say,
to the sublattice $\vec{S}_1$. In this paper we concentrate on the 
contact hyperfine interaction, assuming that $\vec{\sigma}$ is a
nuclear spin. Then, choosing the easy axis/easy plane magnetic anizotropy, 
we write the starting Hamiltonian as follows:
\begin{equation} 
\hat{H} = J \vec{S}_1 \cdot \vec{S}_2 + U(\vec{S}_1) + U(\vec{S}_2) 
+ {\omega_o \over 2 S_1} \vec{S}_1 \cdot \vec{\sigma}, \;\; U(\vec{S}) = 
- K_{\parallel} S_{z}^2 + K_{\perp} S_{y}^2 - \gamma_e \vec{H}_o \vec{S},
\label{e1}
\end{equation}
where $\omega_o$ is the  hyperfine frequency, $K_{\parallel} > 0$ and 
$K_{\perp} > 0$ are magnetic anizotropy constants, and we assume that
$J,\: K_{\perp} \gg K_{\parallel}$, 
$\gamma_e S_{1,2} \mid \vec {H}_o \mid \ll 2K_{\parallel} S_{1,2}^2 $, 
$\omega_o \ll 2K_{\parallel} S_{1,2}^2 $. (We do not confine ourselves by 
$J \gg K_{\perp}$, although in practice this limit realizes very often). 
Assuming $S_1+S_2 \gg 1$, we will use the {\it semiclassical} description. 

In the low-energy dynamics the dominant  
contribution to the transition amplitude 
between the two semiclassical energy 
minima comes  from trajectories with almost {\it antiparallel} $\vec{S}_1$ 
and $\vec{S}_2$. That is $\vec{S}_1$ and $\vec{S}_2$ tunnel simultaneously, 
with only small fluctuations about the comon axis, and the 
tunneling is described by the Neel vector $\vec{N}=\vec{S}_1-\vec{S}_2$ with 
almost constant modulus. As we assume in the  general case that 
$S_1 \ne S_2$, an excess spin $S=S_1-S_2$ will tunnel together with the Neel 
vector (ferrimagnetic case). 

Let us introduce the spherical angles $\theta_1$, $\theta_2$, $\phi_1$, 
$\phi_2$ to describe the orientation of $\vec{S}_1$ and   $\vec{S}_2$. 
We then write  $\theta_2=\pi-\theta_1-\epsilon_{\theta}$,
$\phi_2=\pi+\phi_1+\epsilon_{\phi}$ with  $\vert \epsilon_{\theta} \vert,
\vert \epsilon_{\phi} \vert \ll 1$ (see [4]). Also, we choose as the 
basis set the states
characterized by the opposite orientations of Neel vector (along "easy axis" 
$\vec{z}$), $ \vert \Downarrow \rangle, \vert \Uparrow \rangle$. Then we can 
write the transition amplitude as the path integral:
\begin{equation}
\Gamma_{ \alpha \beta }(t) =
\int_{\beta }^{\alpha} {\cal D} \{ \theta_1,\theta_2,\phi_1,\phi_2 \}
\exp \left\{ -\int_0^t d \tau [ {\cal L}_o (\tau ) + \delta {\cal L}_H(\tau)
+ \delta {\cal L}_{\sigma}(\tau) ] \right\} \;,
\label{e2}
\end{equation}
where ${\cal L}_o(\tau)$ - Lagrangian of the non-interacting problem, 
$\delta {\cal L}_H(\tau)$ describing the weak magnetic field contribution 
and $\delta {\cal L}_{\sigma}(\tau)$ describing the interactions with the 
nuclear spins (all the Lagrangians are in the Euclidean form and 
$\alpha,\beta=\vert \Downarrow \rangle, \vert \Uparrow \rangle$).

In our problem there are two time scales. $\Omega_o^{-1}$ corresponds to the 
transition time between the two semiclassical minima. The second time scale 
is defined by the value of the tunneling amplitude $\Delta_o$ (the tunneling 
splitting is equal to $4 \Delta_o \cos \pi S$). The relation between the 
transition amplitude (\ref{e2}) and the non-diagonal part of the effective 
Hamiltonian is given by [15] (for the real-time axis):
\begin{equation}
\left( H_{eff}^{ND} (\vec{\tau }) \right)_{\alpha \beta  } = {i \over t}
\Gamma^o_{\alpha \beta  }(t)\;; \;\;\;\; (\alpha \ne \beta \;, \;\;
\Omega_o^{-1} \ll t \ll \Delta_o^{-1}) \;.
\label{e12}
\end{equation}

\bigskip

\begin{center}
{\bf B. SEMICLASSICAL SOLUTION} 
\end{center}

Let us find first an extremal trajectory for 
$\theta_1(\tau)$ and 
$\phi_1(\tau)$ for the non-interacting problem when $|\vec{H}_o|=0$ 
and $\omega_o=0$. The Lagrangian ${\cal L}_o$ has the following form: 
\begin{eqnarray}
{\cal L}_o & = & J S_1 S_2 (\sin\theta_1 \sin\theta_2 \cos(\phi_1-\phi_2) 
+ \cos\theta_1 \cos\theta_2 + 1) + {\cal L}_{\vec{S}_1}^o 
+ {\cal L}_{\vec{S}_2}^o, \nonumber \\ 
{\cal L}_{\vec{S}}^o & = & 
- i S \phi \dot{\theta} \sin\theta 
+ K_{\parallel} S^2 \sin^2 \theta 
+ K_{\perp} S^2 \sin^2 \theta \sin^2 \phi. 
\label{e3}
\end{eqnarray}
Since $\vec{y}$-axis is a "hard" one, $\phi$ will weakly fluctuate around
$\phi=0$ or $\phi=\pi$. These values of $\phi$ correspond to two possible 
trajectories connecting two semiclassical minima: clockwise and anticlockwise.
Performing the Gaussian integration over three "fast" variables 
($\phi$, $\epsilon_{\theta}$, $\epsilon_{\phi}$) we get (omitting all the
terms which do not contribute to the equation of motion): 
\begin{eqnarray}
{\cal L}_o (\theta) & = & {{\cal M} \over 2 } \dot{\theta}^2
+ {S_2^2 \over 2 \tilde{J}} \dot{\phi}^2 \sin^2 \theta
+ \tilde{K}_{\parallel} \sin^2 \theta, \;\;\;
\tilde{z} = \tilde{K}_{\perp} - {2 K_{\perp}^2 S_2^4  \over 
\tilde{J} + 2 K_{\perp} S_2^2 }, \; \nonumber \\
{\cal M} & = & {S^2 \over 2 \tilde{z}} 
+ {2 S S_2^3 K_{\perp}  \over \tilde{z} (\tilde{J}+2 K_{\perp} S_2^2) }
+ {S_2^2  \over \tilde{J}+2 K_{\perp} S_2^2 }
+ {2 S_2^6 K_{\perp}^2  \over \tilde{z} (\tilde{J}+2 K_{\perp} S_2^2)^2 } \;,
\label{e4}
\end{eqnarray}
where $\tilde{K}_{\perp,\parallel} = K_{\perp,\parallel} (S_1^2+S_2^2)$ and 
$\tilde{J} = J S_1 S_2$ (below we will ignore indexes at $\theta$ and $\phi$). 
This Lagrangian results in the equation of motion 
$\dot{\theta}=\Omega_o \sin \theta$, the solution of which is:
\begin{equation}
\sin \theta (\tau) = 1 / \cosh (\Omega_o \tau ), \;\;\;\;
\Omega_o =  (2 \tilde{K}_{\parallel} / {\cal M} )^{1/2}  \;, 
\label{e5}
\end{equation}
Substituting the extremal trajectory 
into ${\cal L}_o$ and integrating over $\tau$, we get for the Euclidean 
action: 
\begin{equation}
A_o^{\eta} = A_o + i \eta \pi S, \;\;\;\; 
A_o = 4 \tilde{K}_{\parallel} / \Omega_o, 
\label{e6}
\end{equation}
where $\eta=\pm$ corresponds to the motion between two semiclassical minima 
in clockwise or anticlockwise direction and $\eta \pi S$ is the Kramers-Haldane
phase. 

\bigskip

\begin{center}
{\bf C. MAGNETIC FIELD} 
\end{center}

Now we introduce a weak magnetic field ($\gamma_e=1$): 
\begin{equation}
\delta {\cal L}_H = \delta {\cal L}_{\vec{S}_1}^H
+ \delta {\cal L}_{\vec{S}_2}^H, \;\; \delta {\cal L}_{\vec{S}}^H = 
-  S ( H_o^x \sin \theta \cos \phi 
+ H_o^y \sin \theta \sin \phi + H_o^z \cos \theta ). 
\label{e7}
\end{equation}
If we integrate ${\cal L}_o + \delta {\cal L}_H$ over $\epsilon_{\theta}, 
\epsilon_{\phi}$, with the excess spin $S$ equal to zero and take the limit 
$J \gg K_{\perp}$, then we will get the well-known expression for the 
Lagrangian of Andreev-Marchenko [17] without the gradient term and the term 
describing weak ferromagnetism of Dzyaloshinskii. But we will proceed in the 
general form. After integrating (\ref{e7}) over fast variables 
$\epsilon_{\theta}, \epsilon_{\phi}$ and $\phi$ (keeping only 
linear in magnetic field terms) we substitute the extremal trajectory 
(\ref{e5}) and further integrate over $\tau$ to get a correction to 
the action (\ref{e6}): 
\begin{eqnarray}
\delta A_H^{\eta} =
&-& i { \eta \pi S_2^2 H_y \over \tilde{J} + 2 K_{\perp} S_2^2 }
\left\{ 1 + { 2 K_{\perp}^2 S_2^4 \over \tilde{z} ( \tilde{J}
+ 2 K_{\perp} S_2^2)} \right\} 
- { \eta \pi H_x \over \Omega_o} \left\{ { K_{\parallel} S_2^3 \over
\tilde{J} } + { K_{\perp} S_2^5 \Omega_o^2 \over \tilde{z} 
(\tilde{J}+2 K_{\perp} S_2^2)^2 } \right\} \; \nonumber \\ 
&-& { \eta \pi S \over \Omega_o } \left\{ H_x + i { S \Omega_o  \over 
2 \tilde{z} } H_y \right\} 
- { \eta \pi S \Omega_o S_2^2 \over 2 \tilde{z} (\tilde{J}+2 K_{\perp} S_2^2) }
\left\{ H_x + i { 4 K_{\perp} S_2 \over \Omega_o } H_y \right\} \;.
\label{e8}
\end{eqnarray}

\bigskip

\begin{center}
{\bf D. NUCLEAR SPINS} 
\end{center}

Let us include now coupling between the  electronic spins 
of the nanoparticle and  nuclear spins. The correction to the Lagrangian 
${\cal L}_o$ from hyperfine interaction has a form: 
\begin{equation}
\delta {\cal L}_{\sigma} = 
{ \omega_o \over 2 } \bigg( \hat{\sigma}_x \sin \theta_1 \cos \phi_1 
+ \hat{\sigma}_y \sin \theta_1 \sin \phi_1 + \hat{\sigma}_z \cos \theta_1 
\bigg), 
\label{e9}
\end{equation}
where $\hat{\sigma}_i$ ($i=x,y,z$) are Pauli matrixes. Performing the same 
operations as in the case of magnetic field we get for the correction to 
the action:
\begin{equation}
\delta A_{\sigma}^{\eta} =
{ \eta \pi \omega_o \over 2 \Omega_o } \bigg( \hat{\sigma}_x
+ i { S_2^3 \Omega_o K_{\perp} \over \tilde{z} (\tilde{J}+2 K_{\perp} S_2^2) }
\hat{\sigma}_y \bigg) 
+ i { \eta \pi \omega_o S \over 4 \tilde{z} } \hat{\sigma}_y \;.
\label{e10}
\end{equation}

\bigskip

\begin{center}
{\bf E. EFFECTIVE HAMILTONIAN} 
\end{center}

Since $\Omega_o^{-1} \ll t \ll \Delta_o^{-1} $, 
we can write the non-diagonal part of the effective Hamiltoniah in the form:
\begin{equation}
H_{eff}^{ND} = {i \over t} \left\{ \hat{\tau }_-
\hat{\Gamma}_{ \Downarrow \Uparrow }(t) +  H.c. \right\} , \;\; 
\Gamma_{ \Downarrow \Uparrow }(t) = 
it \Delta_o \sum_{\eta =\pm } \exp \left\{ -A^{\eta} \right\},
\label{e13}
\end{equation}
where $\hat{\tau}_-$ is the lowering operator in the space of Pauli matrixes 
and the transition amplitude is (see also [14]):
\begin{equation}
\Delta_o = { \Omega_o \over 2} \sqrt{{6 \over \pi} A_o } 
\exp \{ -A_o \} \;.
\label{e11}
\end{equation}

We start constructing the effective Hamiltonian from the antiferromagnetic 
case, when $S_1=S_2$ and only Neel vector undergoes tunneling. 
Keeping only linear in $H_o$ and $\omega_o$ terms in the action 
(for the simplicity we set $H_x=0$) we get:
\begin{eqnarray}
A^{\eta}(0) &=& A_o(0) - i \eta \psi(0) + \eta \alpha(0) \bigg( \hat{\sigma}_x 
+ i \lambda(0) \hat{\sigma}_y \bigg), \; 
\Omega_o(0) = 2 S_2 \sqrt { K_{\parallel} (J + K_{\perp}) }, \nonumber \\ 
A_o(0) &=& 4 S_2 \lambda(0), \; 
\psi(0) = { \pi H_y  \over J + K_{\perp} }, \; 
\alpha(0) = { \pi \omega_o \over 2 \Omega_o(0) }, \; 
\lambda(0) = \sqrt{ K_{\parallel} / (J + K_{\perp}) }. 
\label{e14}
\end{eqnarray}

Substituting (\ref{e14}) into (\ref{e13}), we obtain the non-diagonal part of 
the effective Hamiltonian: 
\begin{equation}
H_{eff}^{ND}(0) = 2 \Delta_o(0) \hat{\tau }_ -  \cos \bigg[ \psi(0)
+ \alpha(0) \bigg( i \hat{\sigma}_x
- \lambda(0) \hat{\sigma}_y \bigg) \bigg] + H.c.. 
\label{e15}
\end{equation}
We do not present here the diagonal (static) part. The general way of 
calculating this value one can find in [15].

It is easy to verify that in opposite limiting case, $S_1 \gg S_2$, 
we will get the effective Hamiltonian describing tunneling of the 
ferromagnetic-grain magnetization: 
\begin{eqnarray}
H_{eff}^{ND}(S_1) &=& 2 \Delta_o(S_1) \hat{\tau }_ -  \cos \bigg[ \pi S_1 
- \psi(S_1) -  \alpha(S_1) \bigg( i \hat{\sigma}_x 
- \lambda(S_1) \hat{\sigma}_y  \bigg) \bigg] + H.c. \;, \nonumber \\ 
\Omega_o(S_1) &=& 2 S_1 \sqrt { K_{\parallel} K_{\perp} }, \;\;
\alpha(S_1) = { \pi \omega_o \over 2 \Omega_o(S_1) }, \;\;
\psi(S_1) = { \pi H_y \over 2 K_{\perp} }, \;\; 
A_o(S_1) = 2 S_1 \lambda(S_1), \;\; 
\lambda(S_1) = \sqrt{{ K_{\parallel} \over K_{\perp} }}, 
\label{e16}
\end{eqnarray}
where $S_1$ is understood as the grain's "giant spin".

Finally, in the case of arbitrary value of $S$ (ferrimagnetic) the Neel
vector tunnels {\it together} with the excess spin. In the limit of 
$\tilde{J} \gg 2 K_{\perp} S_2^2$ the effective Hamiltonian has the following 
form (as before $H_x=0$): 
\begin{eqnarray}
H_{eff}^{ND}(S) &=& 2 \Delta_o(S) \hat{\tau }_ -  \cos \bigg[ \pi S - \psi(S) 
- \alpha(S) \bigg( i \hat{\sigma}_x - \lambda(S) \hat{\sigma}_y \bigg) \bigg] 
+ H.c., \nonumber \\ 
A_o(S) &=& \sqrt{ 4 { K_{\parallel} \over K_{\perp} } \bigg( 
S^2 + S_2^2 { 2 \tilde{K}_{\perp} \over \tilde{J} } 
+ S S_2^3 { 4 K_{\perp} \over \tilde{J} } \bigg) }, \;\; 
\lambda(S) = { \Omega_o(S) \over \tilde{K}_{\perp} } \bigg( {S \over 2} 
+ { K_{\perp} S_2^3 \over \tilde{J} } \bigg), \nonumber \\ 
\psi(S) &=& { \pi A_o(S) H_y \over 2 \Omega_o(S) }, \; 
\alpha(S) = { \pi \omega_o \over 2 \Omega_o(S) }, \; 
\Omega_o(S) = \sqrt{ { 4 \tilde{K}_{\parallel} \tilde{K}_{\perp} \tilde{J} 
\over S^2 \tilde{J} + 2 \tilde{K}_{\perp} S_2^2 + 4 S S_2^3 K_{\perp} } }. 
\label{e17}
\end{eqnarray}

From the analysis of these three Hamiltonians the criterion (on the value of 
the excess spin), corresponding to the switching between the ferromagnetic and 
the antiferromagnetic regimes, follows directly:
\begin{eqnarray}
S &\gg& S_2 \sqrt{ {2 \tilde{K}_{\perp} \over \tilde{J} + 2 K_{\perp} S_2^2} }
\;\; - \;\; ferromagnetic \; regime   \nonumber \\ 
S &\ll& S_2 \sqrt{ {2 \tilde{K}_{\perp} \over \tilde{J} + 2 K_{\perp} S_2^2} }
\;\; - \;\; antiferromagnetic \; regime. 
\label{e18}
\end{eqnarray}

\bigskip

\begin{center}
{\bf III. EXACT DIAGONALIZATION} 
\end{center}

To verify the obtained expressions we use the exact diagonalization 
method in application to the starting Hamiltonian (\ref{e1}) with $H_x=0$. 
Because of the limited space, we present the numerical results only for the 
antiferromagnetic case ($S_1=S_2$). The exact diagonalization procedure is 
described in detail in [15], and therefore we define here only the values 
that are subject to the analysis. 
Expanding (\ref{e15}) into the series up to the linear in $\omega_o$ terms 
we get:
\begin{equation}
H_{eff}^{ND} \approx 
2 \Delta_o(0) \bigg( \cos{\psi(0)} \cdot {\hat{\tau}}_x {\hat{\sigma}}_o 
- \alpha(0) \sin{\psi} \cdot {\hat{\tau}}_y {\hat{\sigma}}_x 
+ \alpha(0) \lambda(0) \sin{\psi} \cdot {\hat{\tau}}_x {\hat{\sigma}}_y 
\bigg). 
\label{e19}
\end{equation} 

An effective Hamiltonian defined by the exact 
diagonalization has the form:
\begin{equation}
H=\sum_{\alpha  ,\beta } C_{\alpha  ,\beta } {\hat {\tau }}_{\alpha }
{\hat {\sigma }}_{\beta } \equiv 
C_{xo} \cdot {\hat {\tau }}_x {\hat {\sigma }}_o + 
C_{xy} \cdot {\hat {\tau }}_x {\hat {\sigma }}_y + 
C_{yx} \cdot {\hat {\tau }}_y {\hat {\sigma }}_x +H_D\;,
\label{e20}
\end{equation}
where $H_D$ is diagonal in the ${\hat {\tau }}_z$-representation.
The coefficients $C_{ij}$ are presented in Figures 1.a, 1.b and 1.c 
(dashed lines) together with the  corresponding analytical values from (\ref{e19}) 
(solid lines) as functions of magnetic field at 
$J=40, K{\perp}=20, K_{\parallel}=1, \omega_o=0.2, S_1=S_2=10,\sigma=1/2$.
In Fig. 1.d the coefficient $\alpha$ extracted from $C_{yx}$ (dashed line) 
is shown together with its analytical analog from (\ref{e19}) (solid line) 
as a function of sublattice spin at $J=60, K_{\perp}=20, K_{\parallel}=1, 
\omega_o=0.2, \sigma=1/2$. This coefficient describes the interaction 
with the spin bath and, hence is the most interesting for the present 
investigation. As one can see from all the Figures, the coincidence 
of the analytical results with the numerical ones is quite satisfactory. 
Some differences between the 
two values for $\alpha$ at small $S_2$ are explained 
by the fact that semiclassical description is far from being valid
in this region.

\bigskip

\begin{center}
{\bf IV. CONCLUSIONS} 
\end{center}

Though the {\it detailed} analysis of all 
the factors effecting the tunneling process is beyond the scope of this paper, 
it is possible to make some qualitative conclusions (even without rigorous
calculations with obtained Hamiltonians) on the spin dynamics 
of the investigated systems: (i) the quantum tunneling effect is really more 
strong in antiferromagnetic nanoparticles than in ferromagnetic ones that can 
be seen even from the fact that $A_o(0) < A_o(S_1)$; 
(ii) the presence of the excess spin strongly effects the quantum tunneling 
process due to the mere fact that it renormalizes $A_o$ and $\Omega_o$ which 
both define the tunneling splitting and the bounce frequency; 
(iii) interaction with the environmental spins is extremely important 
since it can drastically change the tunneling picture down to the full 
suppression of it, as in the case of tunneling of half-integer excess spin or
half-integer "giant spin".

\bigskip

\begin{center}
{\bf ACKNOWLEDGMENTS}
\end{center}

The author is grateful to N. Prokof'ev, P.C.E. Stamp and B. Barbara for 
very helpful discussions. The author also acknowledges the hospitality of
the Louis Neel Laboratory (Grenoble, France), where a part of this work has 
been done. This work was supported by the Russian Foundation for Basic 
Research (grant 97-02-16548).

\newpage

\begin{center}
{\bf FIGURE CAPTIONS }
\end{center}

\figure{{\bf Figure 1 } (a)-(c) The numerical coefficients $C_{ij}$ (dashed 
line) in comparison with the analytical expressions from (\ref{e19}) (solid 
line) as functions of magnetic field $H_y$ at 
$J=40, K{\perp}=20, K_{\parallel}=1, \omega_o=0.2, S_1=S_2=10,\sigma=1/2$.
(d) The numerical (dashed line) and the analytical (\ref{e19}) (solid line) 
dependences of the coefficient $\alpha$ on the value of the sublattice spin 
at $J=40, K{\perp}=20, K_{\parallel}=1, \omega_o=0.2, \sigma=1/2$.}

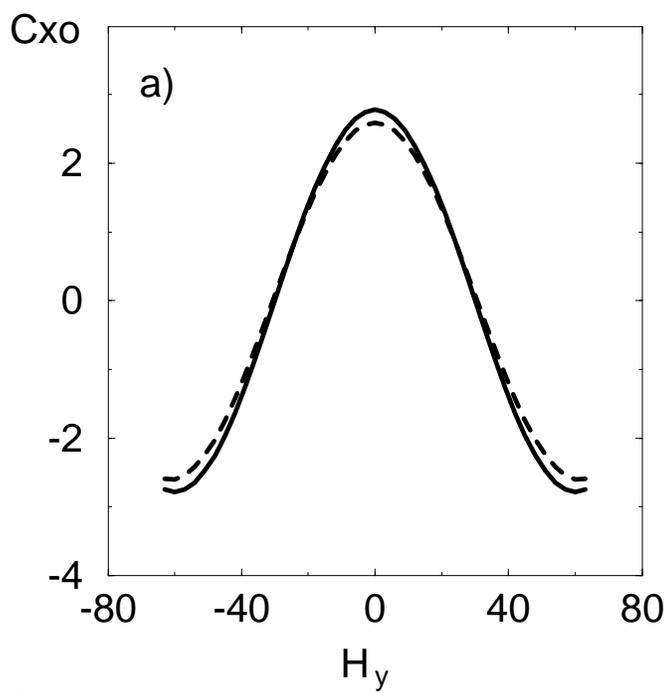
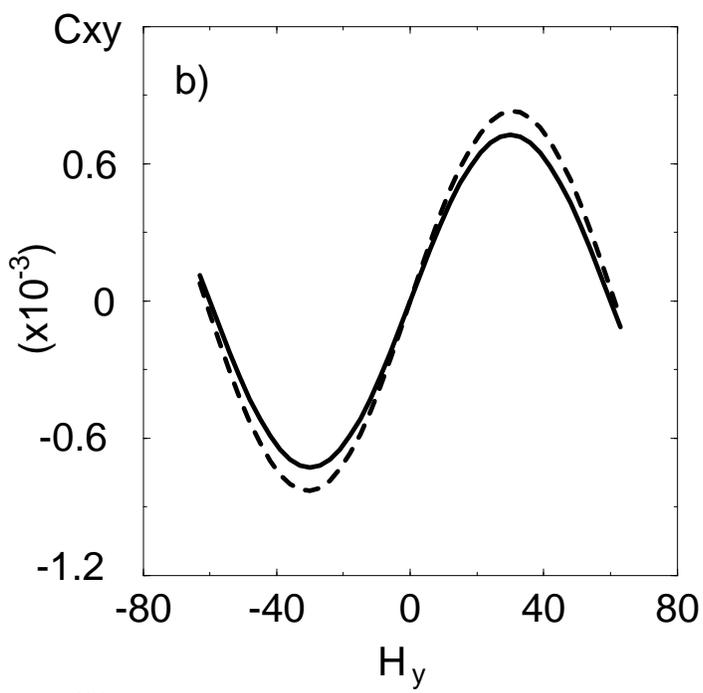
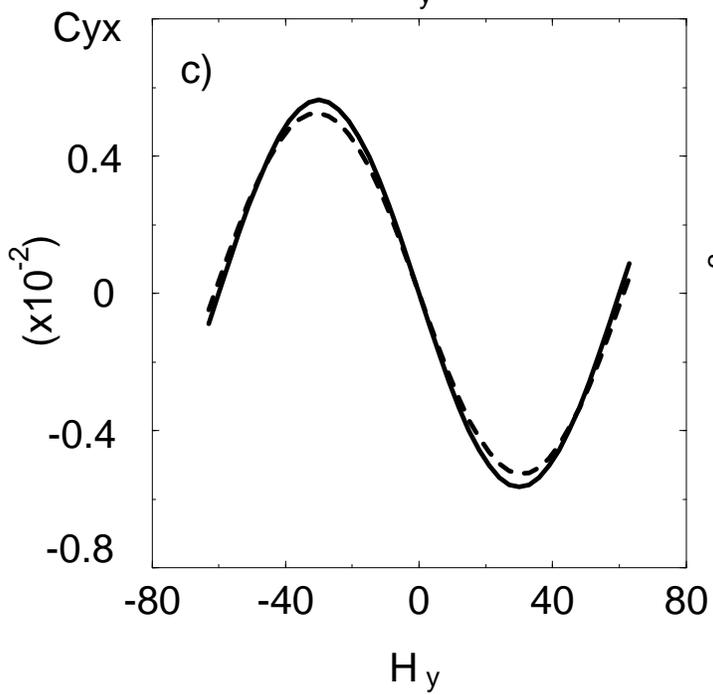
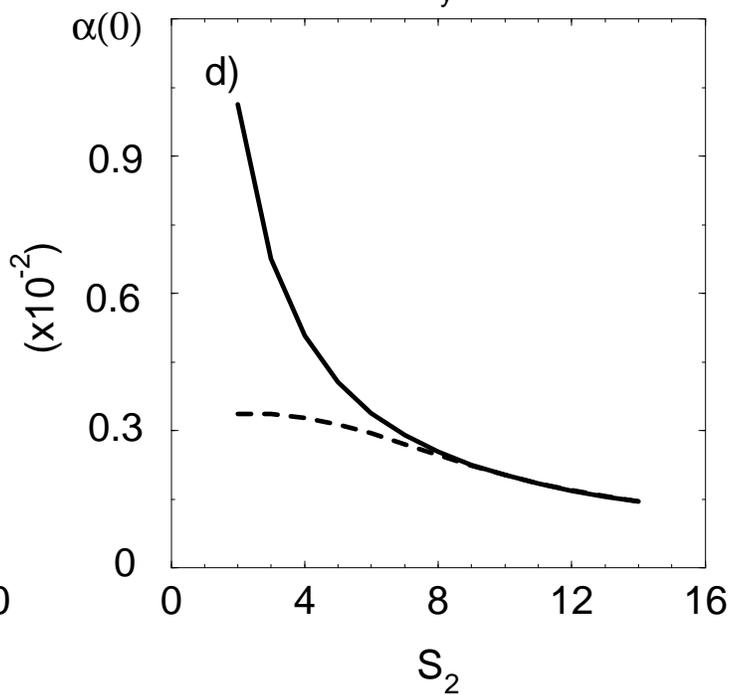
\end{document}